 \documentclass[twocolumn,aps,superscriptaddress,floatfix,prl]{revtex4-1}

 \usepackage{amsmath}
 \usepackage{amssymb}
 \usepackage{xcolor}
 \usepackage{braket}
 \usepackage{dsfont}
 \usepackage{lineno}
 \usepackage{float}
 \usepackage{graphicx}
 \usepackage{mathtools}
 \usepackage{hyperref}
 \usepackage{amsthm}
 \usepackage{verbatim}
 \usepackage{tikz}
 \usepackage{subfigure}
 \usepackage{lipsum}
 \usepackage[height=24cm, width=19cm]{geometry}
 
\begin{document}



\title{Witnessing Environment Induced Topological Phase Transitions via Quantum Monte Carlo and Cluster Perturbation Theory Studies}

\author{F. Pavan}
\affiliation{Dip. di Fisica E. Pancini - Universit\`a di Napoli Federico II - I-80126 Napoli, Italy}

\author{A. de Candia}
\affiliation{SPIN-CNR and Dip. di Fisica E. Pancini - Universit\`a di Napoli Federico II - I-80126 Napoli, Italy}
\affiliation{INFN, Sezione di Napoli - Complesso Universitario di Monte S. Angelo - I-80126 Napoli, Italy}

\author{G. Di Bello}
\affiliation{Dip. di Fisica E. Pancini - Universit\`a di Napoli Federico II - I-80126 Napoli, Italy}

\author{V. Cataudella}
\affiliation{SPIN-CNR and Dip. di Fisica E. Pancini - Universit\`a di Napoli Federico II - I-80126 Napoli, Italy}
\affiliation{INFN, Sezione di Napoli - Complesso Universitario di Monte S. Angelo - I-80126 Napoli, Italy}

\author{N. Nagaosa}
\affiliation{RIKEN Center for Emergent Matter Science (CEMS), Wako, Saitama, 351-0198, Japan}

\author{C.~A. Perroni}
\affiliation{SPIN-CNR and Dip. di Fisica E. Pancini - Universit\`a di Napoli Federico II - I-80126 Napoli, Italy}
\affiliation{INFN, Sezione di Napoli - Complesso Universitario di Monte S. Angelo - I-80126 Napoli, Italy}

\author{G. De Filippis}
\affiliation{SPIN-CNR and Dip. di Fisica E. Pancini - Universit\`a di Napoli Federico II - I-80126 Napoli, Italy}
\affiliation{INFN, Sezione di Napoli - Complesso Universitario di Monte S. Angelo - I-80126 Napoli, Italy}

\begin{abstract}

Many-body interactions play a crucial role in quantum topological systems, being able to impact or alter the topological classifications of non-interacting fermion systems. In open quantum systems, where interactions with the environment cause dissipation and decoherence of the fermionic dynamics, the absence of hermiticity in the subsystem Hamiltonian drastically reduces the stability of the topological phases of the corresponding closed systems. Here we investigate the non-perturbative effects induced by the environment on the prototype Su-Schrieffer-Heeger chain coupled to local harmonic oscillator baths through either intra-cell or inter-cell transfer integrals. Despite the common view, this type of coupling, if suitably engineered, can even induce a transition to topological phases. By using a world-line Quantum Monte Carlo technique we determine the phase diagram of the model proving that the bimodality of the probability distribution of the polarization signals the emergence of the topological phase. We show that a qualitative description can be obtained in terms of an approach based on the Cluster Perturbation Theory providing, in particular, a non-Hermitian Hamiltonian for the fermionic subsystem and insights on the dissipative dynamics. 
    
\end{abstract}
\maketitle










\textit{Introduction}.
\label{introduction}
\color{black}
Topological condensed matter physics is a relatively new field of research that has garnered significant interest in recent years \cite{topological1,topological2,topological3,topological4,topological5,topological6,naoto1,naoto2}. It is based on the notion of topology, a branch of mathematics that studies the geometric properties of objects that remain unchanged under continuous
deformations. The topological properties can give rise
to new states of matter that are protected from external disturbances
and exhibit unique and nontrivial characteristics. The
robustness of these states represents a very peculiar feature
among the wide world of quantum systems and fundamental
in the context of quantum technologies. As such there has been
growing interest in many branches of physics like quantum
information and quantum computing \cite{QC1,QC2,QC3,QC4}, ultracold atoms \cite{UC1,UC2,UC3,UC4,UC5,UC6,UC7,UC8,UC9}
and topological photonics \cite{TP1,TP2,TP3,TP4}.

Commonly quantum phase transitions are classified according
to their critical behavior and can be described by the framework
of the Landau-Ginzburg-Wilson theory \cite{sachdev2011}, which includes
concepts like scaling laws and critical exponents. The critical
behaviour is typically associated with a local order parameter
that undergoes a continuous or discontinuous change.
Topological phase transitions, on the other hand, do not fit
within this framework. They can manifest as the appearance or disappearance
of gapless edge or surface states, or the modification of topological
properties, described physically by topological invariants
or global properties of the system, such as the Chern number or the number of edge states. In this context, the internal symmetries (time-reversal, particle-hole, and chiral symmetries) of the bulk Hamiltonian play a central role in the theory, and all possible combinations thereof give rise to the Altland-Zirnbauer (AZ) classification. \cite{AZ}. 

The natural question addressed in the last decade is if interactions can influence the topological phases of matter and alter the topological classifications of noninteracting fermion systems. It has been proved that these influences can manifest in diverse manners. For instance, there are several examples of how interactions between particles can destroy some topological phases. In some cases interactions can even induce novel topological phases that are not possible in non-interacting systems \cite{interactions-topology}. This statement opens the possibility to take advantage from interactions between the particles to tune between different topological phases \cite{Schirò,floquet1,floquet2}. 

There is another way to affect the topological properties: it stems from the interaction between two different fields. A typical example is provided by open quantum systems \cite{open1,open2}. Since perfect isolation of quantum systems is impossible, interaction with the environment plays a crucial role: it can affect the system even in a non-perturbative way \cite{perroni,def1}. The environment is usually described by an infinite set of harmonic quantum oscillators whose frequencies and coupling strengths obey specific distributions. Then the interaction of the system particles with the environment is concerned with the physics of two coupled fields. The exact elimination of the environment degrees of freedom leads to an effective interaction between the particles of the system that is time retarded and leads to a non- Hermitian Hamiltonian (NHH) for the system under consideration, inducing dissipation, loss of coherence and information. At first glance it is expected to be detrimental to preserve topological phases of matter. In this framework, the seek of a theory capable of generalizing topological phase transitions for physical systems described by NHH has been the focus of an extensive research activity \cite{NH1,NH2,NH3,def}. The non-Hermiticity gives rise to a lot of very interesting phenomena, like the skin effect, compared to Hermitian counterpart. On the other hand, most studies of the last decade concerned with the investigation of NHHs describing non interacting particles in the presence of ad-hoc terms simulating the presence of the environment. In general, the spectrum turns out to be complex and the time evolution of the system is clearly non unitary. In other cases the environment is explicitly taken into account through Lindblad master equation approach, that is appropriate in the limit of weak Markovian interaction \cite{cooper}. 

In order to go beyond these approximations and investigate the physics in any coupling regime, we include the environment in the Hamiltonian, that describes the simplest one-dimensional model exhibiting topological features, i.e. the Su-Schrieffer-Heeger (SSH) model: spinless fermions on a lattice with a two site unit cell and one electron per unit cell (half-filling). Electrons are coupled with local bosonic baths in two different ways, i.e. modulating either intra-cell hoppings $v$ or inter-cell hopping $w$. We emphasize that in our description, where the dynamics of the quantum baths is explicitly taken into account, the full Hamiltonian is Hermitian. 
Since, in general, decoherence effect is viewed as an undesirable destructing factor, the main question we want to address is: can the coupling with the environment be tailored such that topology is induced in a trivial insulator? By using a quantum Monte Carlo (QMC) approach, we will prove that, in one of two investigated cases, the fermion-boson interaction turns a trivial phase in a topological insulating phase. 
In particular, we show that, if open boundary conditions are adopted, the probability distribution of the polarization operator displays a bimodal character near the topological phase transition, both in the absence and in the presence of the interaction with the environment, signalling the appearance of edge states. It paves the way for a new approach capable to unveil the emergence of topological quantum phase transitions. Finally we propose a simple description of the static and dynamical properties of the fermionic system by using Cluster Perturbation Theory (CPT) \cite{CPT1,CPT2,gnote} and periodic boundary conditions in the thermodynamic limit. It allows also to derive an effective NHH for the electrons and points out the validity of the bulk-boundary correspondence in the considered model.   

\color{black}
\textit{The Model}. The SSH model \cite{SSH1,SSH2} (see Fig.\ref{fig:1}) describes spinless fermion hopping on a 1D-chain with staggered transfer integral amplitudes. The chain is made up by $N$ unit cells, each one composed by two sites, indicated with the sub-lattice indices $A,B$. The Hamiltonian  is the sum of two contributions $H_{SSH}=H_v+H_w$: 
\begin{equation}
  H_{SSH}=(v\sum^{N}_{n=1}c^{\dagger}_{n,A}c_{n,B} +w\sum_{n=1}^{N-1} c^{\dagger}_{n+1,A}c_{n,B})+h.c..
\label{ssh bare}
\end{equation}
Here $c_{n,\nu}$ ($c^{\dagger}_{n,\nu}$) destroys (creates) an electron in the site $\nu=A,B$ of the $n$-th cell, with $n=1,...,N$, and $v$ ($w$) denotes the intra- (inter-) cell hopping. This simple toy model underlines the main physical features of topological insulators at half filling. When open boundary conditions are adopted, the trivial insulating phase is obtained for $v>w$, whereas the topological phase, characterized by the appearance of gap-less edge states, sets in at $v<w$. On the other hand, in the thermodynamic limit and assuming periodic boundary conditions, the presence of spontaneous polarization is marked by a topological invariant, $\eta$, that assumes a non vanishing value in a discontinuous way. In this letter we investigate the SSH chain in the presence of local heat baths: $H=H_{SSH}+H_{B}+H_{SSH-B}$. Here $H_{B}=\sum_{n} H_{B,n}$ describes the environment, i.e. local baths where each of them is represented as a collection of harmonic oscillators with frequencies $\omega_{\alpha,n}$, $H_{B,n}=\sum_{\alpha}\omega_{\alpha,n}b^{\dagger}_{\alpha,n}b_{\alpha,n}$, 
and $H_{SSH-B}$ denotes the fermion-boson interaction:
\begin{figure}
    \centering
    \includegraphics[scale=0.3]{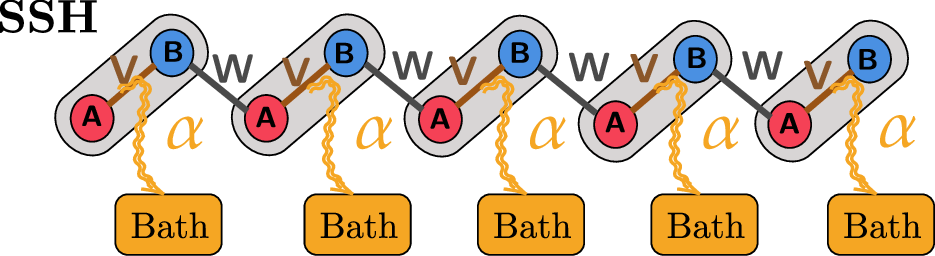}
    \caption{Schematic view of the SSH model in the presence of intra-cell hopping coupled with local bosonic baths.}
    \label{fig:1}
\end{figure}
\begin{equation}
H_{SSH-B}=\sum_{\alpha,n}g_{\alpha,n} (b^{\dagger}_{\alpha,n}+b_{\alpha,n})H_{hop,n}.
\label{int}
\end{equation}
In Eq.(\ref{int}) $H_{hop,n}$ describes the intra- ($c^{\dagger}_{n,A}c_{n,B}+h.c.$) or the inter-cell ($c^{\dagger}_{n+1,A}c_{n,B}+h.c.$) fermionic hopping. We will limit our attention to Ohmic baths, i.e. the spectral density is given by $J(\omega)=\sum_{\alpha} g^2_{\alpha,n}\delta(\omega-\omega_{\alpha,n})=\alpha \Tilde{\omega}\left(\frac{\omega}{\omega_D}\right)e^{-\omega/\omega_D}$ independently on $n$. Here the dimensionless parameter $\alpha$ measures the strength of the coupling, $\omega_D$, Debye frequency, denotes the cutoff frequency, that is assumed to be the largest energy scale, and $\Tilde{\omega}$ is equal to $v$ ($w$) for intra- (inter-) cell couplings between fermions and bosons. In the following we will fix $\omega_D=10 w$ and use units such that $\hbar=a=e=k_B=1$, $a$, $e$ and $k_B$ being the lattice parameter, the electronic charge and Boltzmann constant, respectively. We emphasize that the Hamiltonians considered in this letter, containing two kinds of fermion-boson couplings, fulfill both particle-hole and chiral symmetries, i.e. the symmetries of the bare SSH model are not broken. 

\textit{Open boundary bonditions: QMC approach}. The concept of macroscopic polarization is crucial for describing dielectric media. The polarization operator represents the dipole electrical momentum per unit length and is defined by
$p=\sum_{n=1}^N\left(\frac{2n-1-N}{N-1}\right)\left(c_{n,A}^\dag c_{n,A}+c_{n,B}^\dag c_{n,B}\right)$. It is well known that the dipole momentum, being proportional to the quantum mechanical position operator, is ill defined within periodic boundary conditions. Then, in the first part, we will apply open boundary conditions and use a world-line QMC approach to compute equilibrium values of any quantum observable. Through Suzuki-Trotter decomposition, the imaginary time turns out to be discretized in $2L$ steps of width $\Delta\tau=\frac{\beta}{L}$ ($\beta=1/T$ is the reciprocal of the temperature), and the two terms of the Hamiltonian (\ref{ssh bare}) act on alternate plaquettes in a checkerboard fashion \cite{evertz}. On each site of the square lattice obtained there is a variable $n_{n,\nu,j}=0,1$, where $n=1,\ldots,N$ is the spatial coordinate of the cell, $\nu=A,B$ denotes the site within cell, and $j=1,\ldots,2L$ the temporal coordinate. The occupied sites $n_{n,\nu,j}=1$ form closed lines, that can be considered the world-lines of particles. Particles can hop only on the shaded plaquettes. In the absence of dissipation we use the normal loop algorithm to extract world-line configurations \cite{evertz}. The mean values $\langle H_v\rangle$ and $\langle H_w\rangle$ turn out to be, respectively, $-\beta^{-1}\langle n_{\text{hop}}^{(1)}\rangle$ and $-\beta^{-1}\langle n_{\text{hop}}^{(2)}\rangle$, where $n_{\text{hop}}^{(1,2)}$ are the number of hoppings intra-cell (between $n,A$ and $n,B$) and inter-cell (between $n,B$ and $n+1,A$).
The polarization at a given imaginary time $j$ is $p_j=\sum_{n=1}^N\left(\frac{2n-1-N}{N-1}\right)\left(n_{n,A,j}+n_{n,B,j}\right)$, so that the distribution of polarization $P(p)$ can be computed from the values of $p_j$ at each imaginary time and QMC step.

In the presence of the environment, other variables have to be included in the world-line representation of the system, that is a set of links between two hoppings. We call hopping a plaquette where a particle hops between two sites (within a cell or between two adjacent cells.) Such links represent the emission and absorption of a phonon, and, because the baths are local, the two plaquettes linked must be on the same pair of sites (but at different imaginary times). Each link introduces an additional weight to the configuration, given by $\Tilde{\omega}^{-2}K\left(\frac{\Delta\tau}{2}|j-j^\prime|\right)$, where $1\le j,j^\prime\le 2L$ are the imaginary times of the plaquettes linked, and $K(\tau)=\int\!d\omega\,J(\omega)\frac{\cosh\left[\omega\left(\frac{\beta}{2}-\tau\right)\right]}{\sinh\left(\beta\omega/2\right)}$ ($K$ stems from the exact elimination of the bath degrees of freedom). The QMC dynamics then proceeds in two alternating steps. In the first one, a dynamics of the links is performed: links between hoppings can be attached and removed with appropriate probabilities satisfying the detailed balance. In the second one, the loop algorithm is performed, including the modification that linked hoppings have to be frozen.

\begin{figure}[thb]
    \includegraphics[width=1.01\columnwidth]{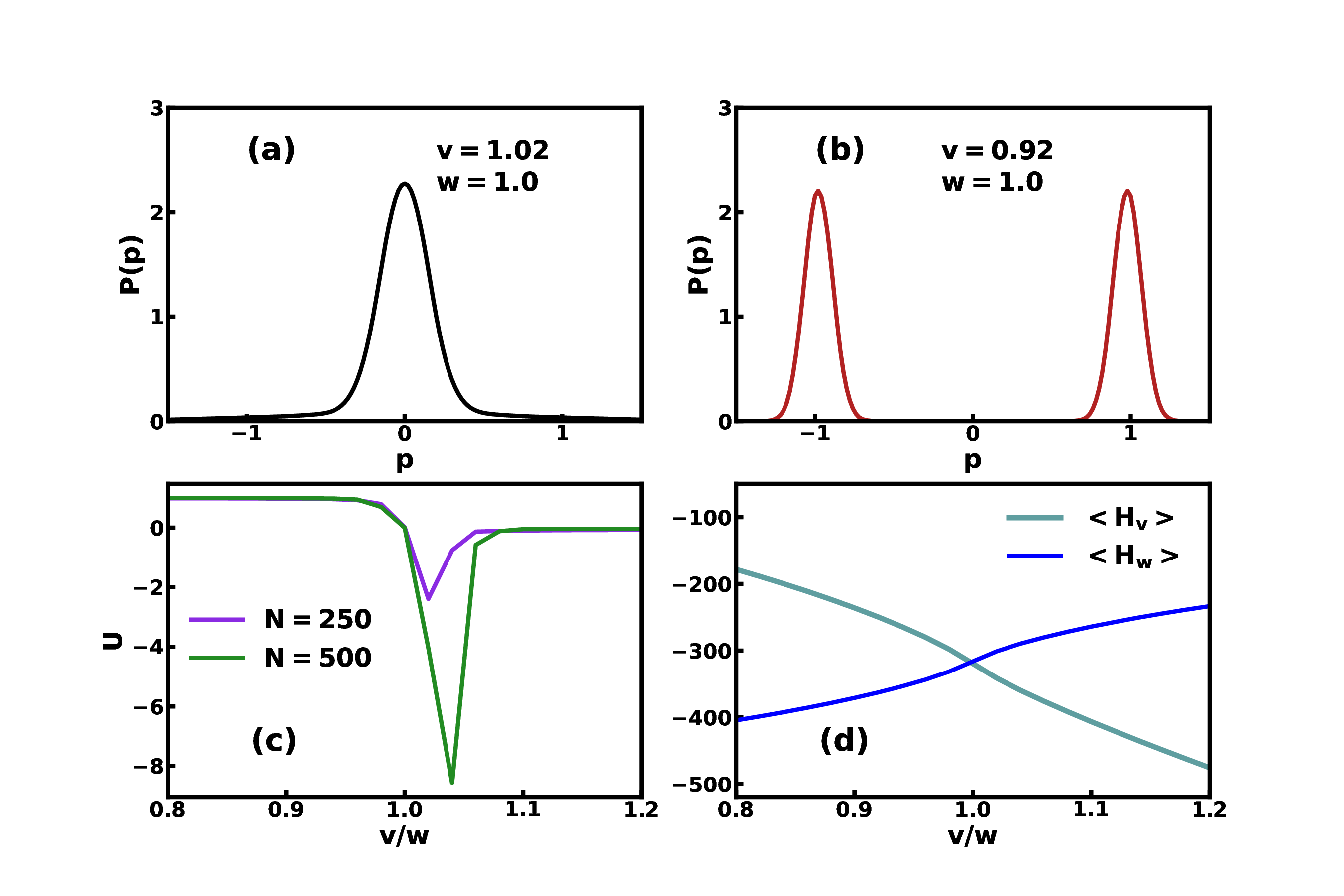}
    \caption{(a) and (b): Polarization probability distribution $P(p)$ in the trivial and topological phases, respectively; (c): Binder parameter vs $v/w$; (d): average values of $H_v$ and $H_w$ vs $v/w$. All plots refer to the non interacting case with $\beta=100$ and $\Delta\tau=0.001$. $N=500$ in (a), (b) and (d).} 
    \label{fig:2}
\end{figure}

Now we prove that the polarization probability distribution, $P(p)$, can be used as marker of topological phase transition. To this aim, first of all we will analyze the behavior of $P(p)$ in the bare SSH where it is well known that $v=w$ represents the border between the trivial and topological insulating phases. In Fig.\ref{fig:2}a, $v>w$, $P(p)$ displays one peak centered at $p=0$. In Fig.\ref{fig:2}b, i.e within the topological phase ($v<w$), $P(p)$ acquires a bimodal character, with two peaks centered at $p=\pm 1$. The presence of two peaks, whose width becomes narrower and narrower by decreasing the temperature $T$and increaing $N$, signals the emergence of the edge states. In Fig.\ref{fig:2}c we plot the Binder parameter, $U$, vs the ratio $v/w$ for different lattice sizes. $U=\frac{1}{2}\left(3-\frac{\langle p^4\rangle}{\langle p^2\rangle^2}\right)$ is the fourth-order cumulant of the distribution $P(p)$ and, in  statistical physics, is frequently used to determine accurately critical points. Similar to standard phase transitions, the plots point out that the different curves cross at $v=w$, changing sign and exhibiting a minimum in the trivial phase, near the threshold, that is the typical behavior within first order discontinuous phase transitions. In Fig.\ref{fig:2}d we plot the average values of the two terms of the Hamiltonian Eq.(\ref{ssh bare}) vs $v/w$ for different lattice sizes. As expected, in the thermodynamic limit, they cross at $v=w$, where the topological quantum phase transition sets in. These observations point out that, in order to identify the occurrence of the topological phase transition, it is possible to use the methods used in the statistical physics to characterize the phase transitions, although the key concepts of Landau classification, i.e. symmetry breaking and local order parameter, are missing in topological phase transitions, where gap-less edge states appear and, at the same time, the bulk properties of the insulators show no changes.  

\begin{figure}[thb]
    \includegraphics[width=1.01\columnwidth]{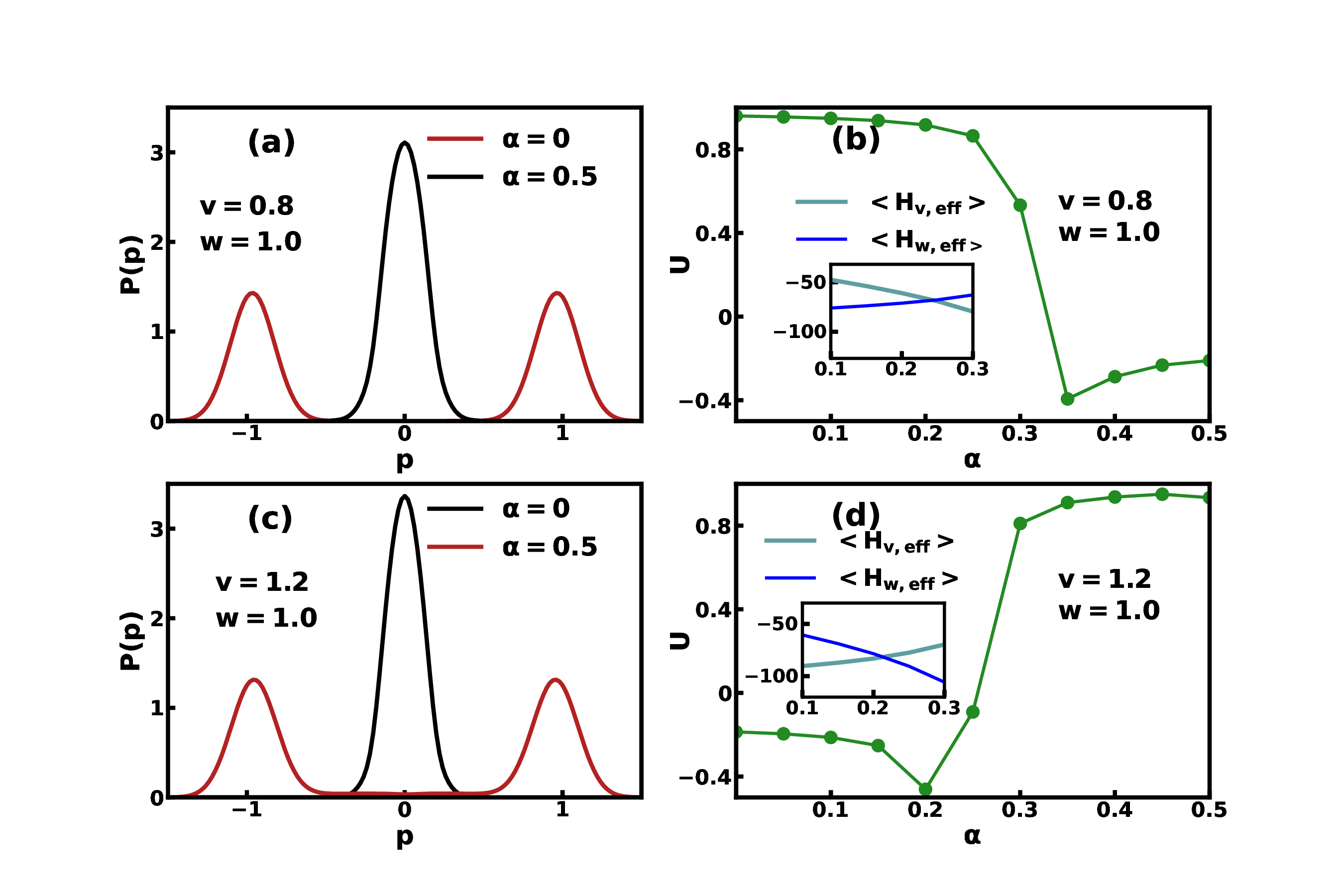}
    \caption{Polarization probability distribution in (a) ((c)) and Binder parameter in (b) ((d)) vs $\alpha$, in the case of intra-cell (inter-cell) hopping coupled with local baths with $N=100$, $\beta=100$ and $\Delta\tau=0.001$. Insets: mean values of $H_{v,eff}$ and $H_{w,eff}$ vs $\alpha$.
    }
    \label{fig:3}
\end{figure}

Now the natural question is: what happens when the interaction with the environment is turned on? In Fig.\ref{fig:3}a and Fig.\ref{fig:3}b the plots of the probability distribution and the Binder parameter show that, if the coupling with local baths affects the intra-cell hopping, environment appears as detrimental for topological states of matter. Starting from a topological insulator, $v<w$, and increasing the strength of the coupling, a phase transition occurs towards a trivial insulator. On the other hand, if the interaction with the bosonic fields acts on the inter-cell transfer integrals, a topological phase transition is induced (see Fig.\ref{fig:3}c and Fig.\ref{fig:3}d). Interestingly, the critical values of the fermion-boson coupling, $\alpha_c$, are in good agreement with the crossing points of the average values of the hopping terms, provided that interaction contributions with the baths are included (see insets). 

\begin{figure}[thb]
    \centering
    \includegraphics[width=1.01\columnwidth]{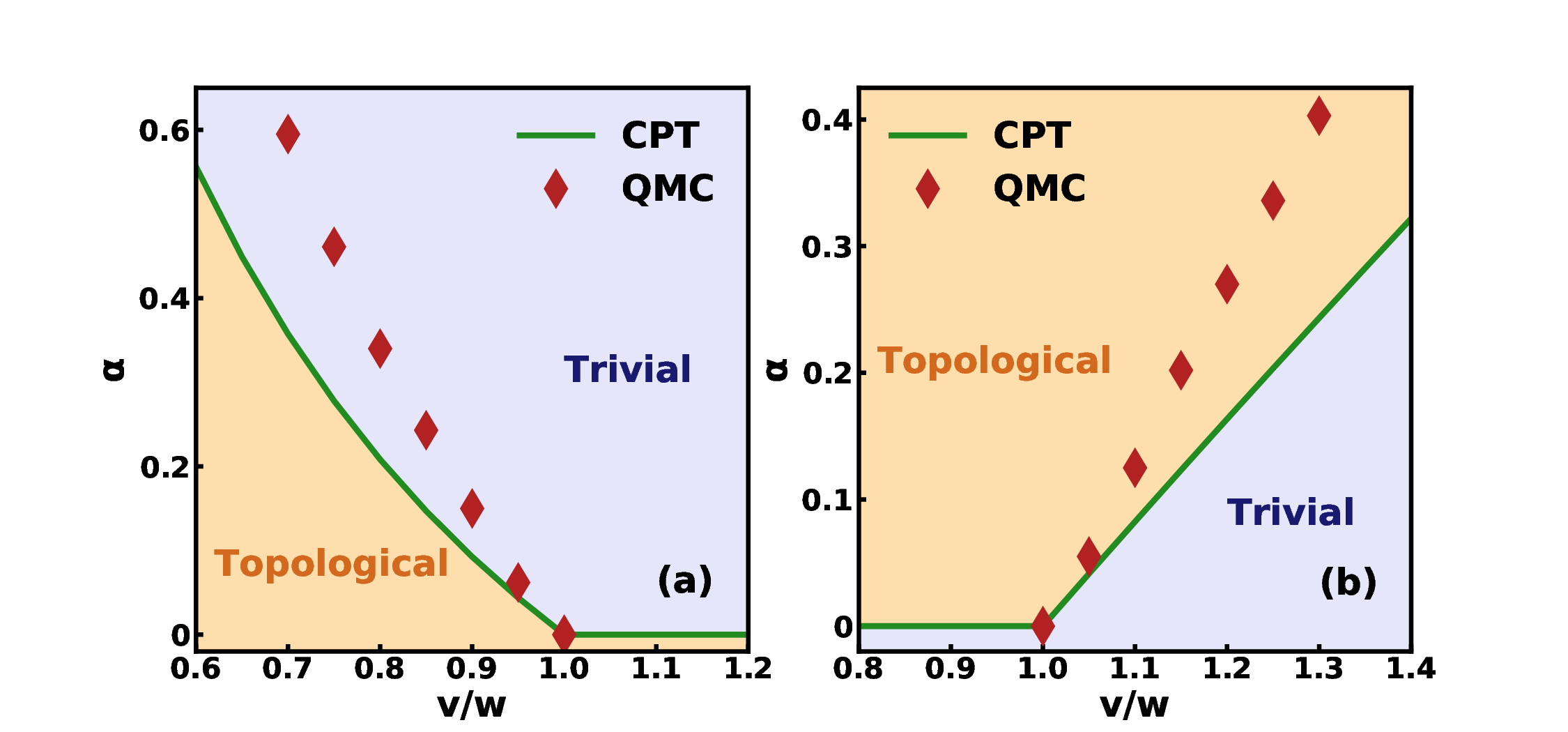}
    \caption{Phase diagram in the case of intra- (inter-) cell hopping coupled with local baths in (a) ((b)). The green line (red diamonds) represents the critical values of the coupling obtained through CPT (QMC) method.}
    \label{fig:4}
\end{figure}

The explanatory phase diagrams, corresponding to intra- and inter-cell fermion-boson couplings, are reported in Fig.\ref{fig:4}a and Fig.\ref{fig:4}b. They clearly show that in the former (latter) case the interactions with local baths is detrimental (favorable) for the emergence of the topology.

\textit{Periodic boundary conditions: CPT}. In order to deal with macroscopic systems, in general, one can assume also periodic boundary conditions. A system of this kind has no surface and all of its properties are by construction bulk ones. In this case the spontaneous polarization is related to the Berry phase and leads to the topological invariant $\eta$. Since the chiral symmetry is preserved, $\eta$ can be evaluated by means of an integral over the Brillouin zone of the Green function at zero frequency \cite{TI1,TI2,TI3,TI4,TI5}: 
\begin{equation}\label{wn}
    \eta=\frac{1}{4\pi i} \int_{-\pi}^{\pi} dk \; Tr \bigg\{\sigma_z G(k,0) \partial_k G^{-1}(k,0)  \bigg\},
\end{equation}  
where $\sigma_z$ is the Pauli matrix. 
CPT \cite{CPT1,CPT2,gnote} is indeed a powerful technique to calculate the Green function in many body systems with local interactions. The basic idea is to divide the lattice into a superlattice of identical clusters. On each of them Green function is solved exactly, whereas the hopping between sites belonging to different clusters is treated perturbatively. Clearly, the validity of the approach improves by increasing the size of the cluster. In this letter we will consider a two site cluster, whose form depends on which case we have been considering. If the intra- (inter-) cell hoppings $v$ ($w$) are coupled with the local baths, the two sites of the cluster belong to the same (different) cell of the SSH chain. In this way, one obtains four mixed Green functions, $G_{i,j}(k,\omega)$ of the interacting system (the indexes $i$ and $j$ can assume two possible values corresponding to the two sites of the cluster). These functions allow us to calculate the topological invariant via Eq.(\ref{wn}) disclosing the emergence of topological features. There are also other Green functions of particular physical interest: $G_{+,+}(k,\omega)$, $G_{-,-}(k,\omega)$, $G_{-,+}(k,\omega)$ and $G_{+,-}(k,\omega)$ corresponding to the quasiparticle operators $\gamma_{k,+},\gamma_{k,-}$ of the bare SSH model, in terms of which the Hamiltonian assumes the diagonal form: $H_{SSH}=\sum_{k} (E_{k,+}\gamma^\dagger_{k,+}\gamma_{k,+}+E_{k,-}\gamma^\dagger_{k,-}\gamma_{k,-})$. They provide detailed pieces of information on the renormalized band structure as well as spectral functions, $A(k,\omega)=-\Im{G(k,\omega)}/\pi$. Clearly, in the absence of fermion-boson coupling, the off diagonal Green functions, $G_{-,+}(k,\omega)$ and $G_{+,-}(k,\omega)$, will vanish, whereas the spectral functions $A_{-,-}(k,\omega)$ and $A_{+,+}(k,\omega)$, corresponding to the diagonal Green function $G_{-,-}(k,\omega)$ and $G_{+,+}(k,\omega)$, are delta functions centered at $E_{k,-}$ and $E_{k,+}$, respectively. 
\begin{figure}[h]
    \centering
    \includegraphics[width=1.01\columnwidth]{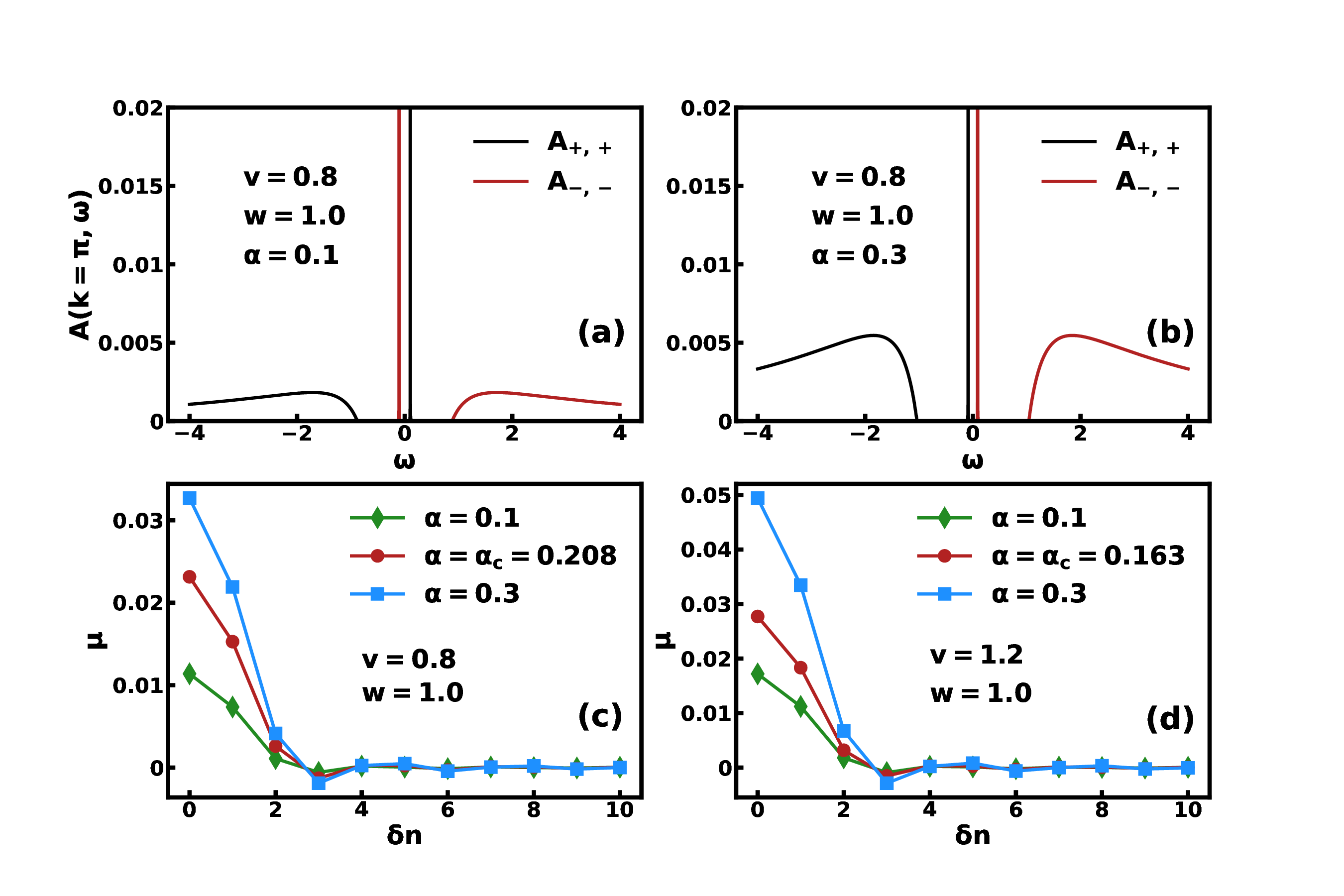}
    \caption{(a) and (b): Spectral Functions $A_{+,+}(k,\omega)$ and $A_{-,-}(k,\omega)$ at $k=\pi$ and couplings below (a) and above (b) the critical value $\alpha_c$, in the case of intra-cell hopping coupled with local baths. (c) and (d): effective sub-lattice hopping $\mu$ vs cell distance $\delta n$ for different values of the polaronic interaction (with $0.1<\alpha_c<0.3$), for intra- and inter-cell fermion-boson couplings.}
    \label{fig:5}
\end{figure}
In Fig.\ref{fig:5}a and Fig.\ref{fig:5}b we plot the spectral weight functions $A_{-,-}(k,\omega)$ and $A_{+,+}(k,\omega)$ at $k=\pi$ for the two considered models. In both cases, by increasing the strength of the coupling with local heat baths, the two main peaks become closer and closer and, at $\alpha=\alpha_c$, become coincident. For values $\alpha>\alpha_c$, the two peaks swap: a closure and reopening of the gap takes place. It is also evident the presence of tails in the spectral functions due to interactions, that is associated with a corresponding reduction of the quasiparticle spectral weight. We emphasize that the critical values of $\alpha$, determined by observing the discontinuous behavior of the topological invariant, are in good agreement with those obtained via QMC technique: the agreement is excellent at small values of $\alpha_c$ and becomes qualitative at high values of $\alpha_c$, due to the small size of the cluster used.

Now we want to prove that it is possible to describe the fermionic subsystem  in terms of an effective NHH (see SM for details \cite{gnote}). The first step is the determination of the poles of the $2 \times 2$ matrix $G_{\pm,\pm}(k,\omega)$ for any fixed value of $k$. They will form two energy bands: $-\Tilde{E}_k-i\delta_k$ and $\Tilde{E}_k-i\delta_k$ with $\delta_k \ge 0$, i.e. the poles are located in the lower half plane. The effective Hamiltonian assumes the following form: 
\begin{equation}\label{ham eff}
H(k)=
\begin{bmatrix} 
\Tilde{E}_k-i\delta_k & 0\\
0& -\Tilde{E}_k-i\delta_k
\end{bmatrix}.
\end{equation}
By considering the inverse of the unitary transformation that allowed us to perform the exact diagonalization of the bare SSH model, it is possible to express this Hamiltonian in terms of the creation and annihilation operators associated to Wannier function within the real space:  $H_{eff}=H_1+H_2+H_3$, where:
\begin{equation}\notag
    H_1=-i\sum_{n,\delta n} \mu(\delta n)\bigg[c^{\dagger}_{A,n+\delta n}c_{A,n}+c^{\dagger}_{B,n+\delta n}c_{B,n} +h.c. \bigg],
\end{equation}
\begin{equation}\notag
    H_2=\sum_{n,\delta n} \Tilde{v}(\delta n) \bigg[c^{\dagger}_{A,n+\delta n}c_{B,n}+c^{\dagger}_{B,n }c_{A,n+\delta n} \bigg],
\end{equation}
and 
\begin{equation}\notag
    H_3=\sum_{n,\delta n} \Tilde{w}(\delta n) \bigg[c^{\dagger}_{A,n+\delta n}c_{B,n}+c^{\dagger}_{B,n }c_{A,n+\delta n} \bigg],
\end{equation}
with $\mu(\delta n)=\frac{1}{2\pi} \int_{0}^{\pi} dk \delta_k \cos[k\delta n]$, $\Tilde{v}(\delta n)=\frac{v}{\pi} \int_{0}^{\pi} dk \frac{\Tilde{E}_k}{E_k} \cos[k \delta n ]$ and $\Tilde{w}(\delta n)=\frac{w}{\pi} \int_{0}^{\pi} dk \frac{\Tilde{E}_k}{E_k} \cos[k (\delta n -1) ]$.

It is evident that: i) $H_1$ (antihermitian operator) describes an effective hopping of the fermions within the same sublattice (this contribution vanishes in the closed system); ii) $H_2$ and $H_3$ (hermitian operators) represent the effective hopping between two sites, each of them being located in one of the two sublattices, whose distance is $\delta n$. Clearly in the absence of interaction, $\Tilde{E}_k=E_k$, $\delta_k=0$ and  the only non-vanishing hoppings are $\Tilde{v}(0)=v$ and $\Tilde{w}(1)=w$, i.e. $H_{eff}$ reduces itself to the bare SSH model. The plots in Fig.\ref{fig:5}c and Fig.\ref{fig:5}d show that, independently of the kind of coupling with local baths, by increasing the strength of the interaction, the quantity $\mu(\delta n)$, that is the main marker of the non-Hermiticity of the subsystem Hamiltonian, becomes larger and larger \cite{hnote}.  

\textit{Conclusions}. By using world line QMC approach and CPT, we investigated, within the SSH model, the effects on the topological quantum phase transitions induced by two different couplings with Ohmic heat local baths, involving either intra-cell or inter-cell hoppings. We proved that whereas the interaction related to intra-cell integral transfer turns out to be detrimental for the topology, the one involving inter-cell hopping can favour the emergence of topological features. Furthermore we demonstrated that: 1) by using open boundary conditions, the bimodality of the polarization probability distribution can be used as marker of the topological phase transition occurrence; 2) CPT provides an effective NHH for the fermionic subsystem, where the main source of the non Hermiticity is given by a pure imaginary transfer integral within the same sublattice.   

\section{acknoledgements} G.D.F. acknowledges financial support from 376 PNRR MUR Project No. PE0000023-NQSTI; N.N. is supported by JST CREST Grant Number JPMJCR1874, Japan.

\bibliography{lib}{}

\begin{thebibliography}{52}
\bibitem [{hno()}]{hnote}%
  \BibitemOpen
  \href@noop {} {}\bibinfo {note} {In general, it is possible to define two
  effective Hamiltonians for the subsystem \cite{TI5}: an Hermitian topological
  Hamiltonian, stemming from Green function evaluated at zero frequency (useful
  to calculate the topological invariant through Eq.3), and a non Hermitian
  Hamiltonian by using the poles of the Green function, i.e. by calculating the
  Green function at no vanishing frequency in the half lower complex plane. In
  general, the latter fails in calculating topological invariants, but it
  captures the quasiparticle physics. In our case, this Hamiltonian correctly
  reproduces also the topological invariant. The most important quantity is the
  renormalized energy: $\Tilde{E_k}$: at $k=\pi$, it becomes zero at the
  critical value of $\alpha$ determining the closure and reopening of the
  gap.}\BibitemShut {Stop}%
\end{thebibliography}
\end{document}